\documentclass[twocolumn]{aastex62}

\usepackage{amsmath}
\usepackage{graphicx}

\usepackage{epsfig}
\usepackage{natbib}
\usepackage{wasysym}
\usepackage{lipsum}
\usepackage{mathrsfs}
\usepackage{float}
\usepackage{rotating}
\usepackage{epstopdf}
\usepackage[graphicx]{realboxes}
\usepackage{array}
\newcolumntype{P}[1]{>{\centering\arraybackslash}p{#1}}
\newcolumntype{M}[1]{>{\centering\arraybackslash}m{#1}}
\newcounter{chem}
\newcounter{temp}
\usepackage{lineno}

\providecommand{\e}[1]{\ensuremath{\times 10^{#1}}}

\usepackage{color}

\usepackage{gensymb}

\usepackage{filecontents}
\usepackage{filecontents}

\begin{document}

\shorttitle{Effects of Libration on Exoplanet Climate}
\shortauthors{Chen, H., et al.}

\title{{\bf  Sporadic Spin-Orbit Variations in Compact Multi-planet Systems \\and their Influence on Exoplanet Climate}}

\author[0000-0003-1995-1351]{Howard Chen}

\affil{Planetary Environments Laboratory, NASA Goddard Space Flight Center,  Greenbelt, MD 20771, USA }
\affil{GSFC Sellers Exoplanet Environments Collaboration, NASA Goddard Space Flight Center,  USA}
\affil{Department of Aerospace, Physics, and Space Sciences, Florida Institute of Technology, Melbourne, FL}

\author[0000-0003-1995-1351]{Gongjie Li}

\affil{School of Physics, Georgia Institute of Technology, Atlanta, GA 30332, USA}

\author[0000-0003-1995-1351]{Adiv Paradise}

\affil{Department of Astronomy and Astrophysics, University of Toronto, Ontario, Canada }

\author[0000-0003-1995-1351]{Ravi K. Kopparapu}

\affil{Planetary Environments Laboratory, NASA Goddard Space Flight Center,  Greenbelt, MD 20771, USA }
\affil{GSFC Sellers Exoplanet Environments Collaboration, NASA Goddard Space Flight Center,  USA}

\begin{abstract}
Climate modeling has shown that tidally influenced terrestrial exoplanets, particularly those orbiting M-dwarfs, have unique atmospheric dynamics and surface conditions that may enhance their likelihood to host viable habitats.
However, sporadic libration and rotation induced by planetary interactions, such as that due to mean motion resonances (MMRs) in compact planetary systems may destabilize attendant exoplanets away from synchronized states (or 1:1 spin-orbit ratio). Here, we use a three-dimensional N-Rigid-Body integrator and an intermediately-complex general circulation model to simulate the evolving climates of TRAPPIST-1 e and f with different orbital and spin evolution pathways. Planet f perturbed by MMR effects with chaotic spin-variations are colder and dryer compared to their synchronized counterparts due to the zonal drift of the substellar point away from open ocean basins of their initial eyeball states. On the other hand, the differences between perturbed and synchronized planet e are minor due to higher instellation, warmer surfaces, and reduced climate hysteresis. This is the first study to incorporate the time-dependent outcomes of direct gravitational N-Rigid-Body simulations into 3D climate modeling of extrasolar planets and our results show that planets at the outer edge of the habitable zones in compact multiplanet systems are vulnerable to rapid global glaciations. In the absence of external mechanisms such as orbital forcing or tidal heating, these planets could be trapped in permanent snowball states.   

\end{abstract}


\section{Introduction} 
\label{sec:intro}

A variety of future observatories are poised to definitively reveal the environmental characteristics of small rocky worlds \citep{Kane+21book}.  Gas-phase species such as  H$_2$O and CO$_2$ have already been detected on potentially water-rich and gaseous exoplanets \citep{KreibergEt2014NAT,BennekeEt2019ApJL,Swain+21,Edwards+21,Fu+22}. For truly terrestrial worlds such as TRAPPIST-1 e and f, simulated detections suggest spectral features of of CO$_2$, CH$_4$, and N$_2$O features at the 4.3, 3.3, and 8.5 $\mu$m bands will be feasible within 30 transits for near-term observations by the JWST (e.g., \citealt{FauchezEt2019ApJ,Lustig-Yaeger+19,Wunderlich+20,Kaltenegger+Lin21,Lustig-Yaeger+22,Mikal-Evans+2022}). In the next few years, more potentially habitable planets will be discovered, with the catalog of these systems to date on order of hundreds and growing; models of various complexities and heritages are just on the rise and will quantitatively evaluate their inhabitance or habitability \citep{Mendez+20,Fauchez+21,Wordsworth+21,Cooke+2022a}.  

Because of the small orbital separation (semi-major axes) of worlds around M-dwarfs, theory suggests that they will be locked in a synchronized state due to strong tidal forces from the host star. As such, the majority of previous work using three-dimensional (3D) general circulation models (GCMs) M-dwarf planets assumed  1:1 spin orbit rotation ratios for planets with short ($\la50$ days) orbital periods (e.g., \citealt{WayEt2016GRL,KopparapuEt2017ApJ,YangEt2019ApJ,ChenEt2019ApJ,Joshi+2020,Guzewich+20,Lefevre+2021,Cohen+22MNRAS,Hammond+Lewis21,Wolf+2022,Sergeev+22,Braam+22}). 
This assumption is acceptable  due to the computational expense and time-consuming nature of many state-of-the-science GCMs. Slowly- and synchronously-rotating exoplanets are fascinating model environments to study unique dynamical features and climate regimes, but there are many means by which their orbits would depart from completely synchronized states. For instance, orbital scattering, merging events, thermal tides, and secular perturbations could drive planets into higher order spin-orbit resonances (SORs) including 6:1. 2:1, and 3:2 \citep{Leconte+2015,Renaud+21}. Using the 3D climate model  Laboratoire de Météorologie Dynamique (LMD/PCM), previous publications have found that the stability of tidally influenced exoplanets increase with asynchronus rotation assumptions (e.g., 3:2 spin-orbit ratio; \citealt{Turbet+2018AA}). \citet{DelGenioEt2019AsBio} however, used the Resolving Orbital and Climate Keys of Earth and Extraterrestrial Environments with Dynamic (ROCKE-3D) model and found that the same resonance case led the lowest global liquid water fractions amongst all their simulations. \citet{YangEt2013ApJL} showed, using the National Center for Atmospheric Research climate model Community Climate System Model (CCSM), that the more rapid rotation rates of 6:1 and 2:1 planets allow the breakup of the dayside stationary cloud decks, which then destabilises the energy balance and push the planets into runaway states close to the IHZ.  Another study, \citet{Colose+2021}, found that in the absence of internal heating, the  IHZ limit is only weakly sensitive to the planet's spin-orbit resonant state. For tidally heated planets however, their results show vastly different evolved climates depending on their assumed resonant state.

The orbital evolution of the Sun-Earth system has an intimate relationship with the observed paleoclimate record, also known as Milankovich cycles. From ice core samples \citep{Kawamura+2007},  N-body simulations \citep{Laskar+2004}, Mars' polar ice caps (e.g., \citep{Toon+1980}, and climate modeling (e.g., \citep{Spiegel+2010,Deitrick+18a,Deitrick+2018b}, for example,  it has been found that the orbital changes on 10-10,000 yr timescale have been crucial in understanding the variability in Earth's climate throughout history.

Along a similar vein, the net orbital motions of distant exoplanets, their host stars, and neighboring rocky bodies exert strong forcings onto their transient and mean climates\footnote{Tidal modulations may also be important for other reasons, for instance, by influencing the stability of planetary orbits \citep{Lingam+22}.}
Planetary rotation, for instance, governs cloud distribution, mean cloud cover, and exoplanetary albedo \citep{YangEt2013ApJL,WayEt2016GRL,Jansen+19,He+22climate}, and can strongly affect  boundaries of the inner edges of the habitable zone \citep{Yang+14}.
Others have found that Earth-like exoplanets can maintain clement temperature at large obliquities \citep{Kilic+17,He+22climate}, avoid global glaciation at lower stellar fluxes due to tidal heating \citep{Colose+19}, and even bolster the oxygenation of exo-ecospheres \citep{Barnett+22}.
Varying eccentricity may also have potentially observable climatic effects  \citep{Way+2017}, and large eccentricity combined with obliquity could instigate ice sheet melting during colder seasons \citep{Shields+16}.  \citet{Vervoort+22} showed how the presence of giant planet induced precession cycles can influence the fractional habitability of planets with Earth-similar atmospheres. Using an energy balance model, \citep{Quarles+22} found that large obliquity
variations (i.e., $>55\degree$ ) can lead to dynamical transitions to snowball stages due to the large thermal inertia of the ice belt.

Future detection and observational measurements will be biased towards short period planets.
As such, compact systems (e.g., multiple planets with $P < 50$ d) of planets close to the host star serve as blueprints for understanding the diversity of exoplanet system architectures \citep{Kane+13AJ,Tamayo+20PNAS}. Owing to the proximity of planets in these systems, mutual gravitational interactions between neighboring planets can cause spin-axis dynamics that may be crucial from  observational and habitability standpoints. For instance, spin-orbit coupling induced orbital precession can lead to detectable transit-timing variations (TTVs; \citealt{Bolmont+20, Chen+21ApJ}), and compete against tidal forces from the host star \citep{Vinson+2019}. To our knowledge, none have previously examined the climate impacts of  chaotic spin-variations due to planet-planet interactions, especially when they dominate over stellar influences. How might these effects modulate a planet's rotational state, amplitude and period of orbit and obliquity cycles, atmospheric evolution, ocean circulation, and surface climate?

In this Letter, we use an N-Rigid-Body integrator and a 3D general circulation model with reduced complexity to evaluate time evolving climates of TRAPPIST-1 planets.  In Section~\ref{methods}, we describe the numerical models and assumptions. In Section~\ref{results}. we present the results of our N-Rigid-Body simulation package and 3D climate model. In Section~\ref{discussion}, we discuss the caveats, implications, and observational relevance of this study. We conclude the study in Section~\ref{conclusion}.

\begin{figure}[t] 
\begin{center}
\includegraphics[width=1.\columnwidth]{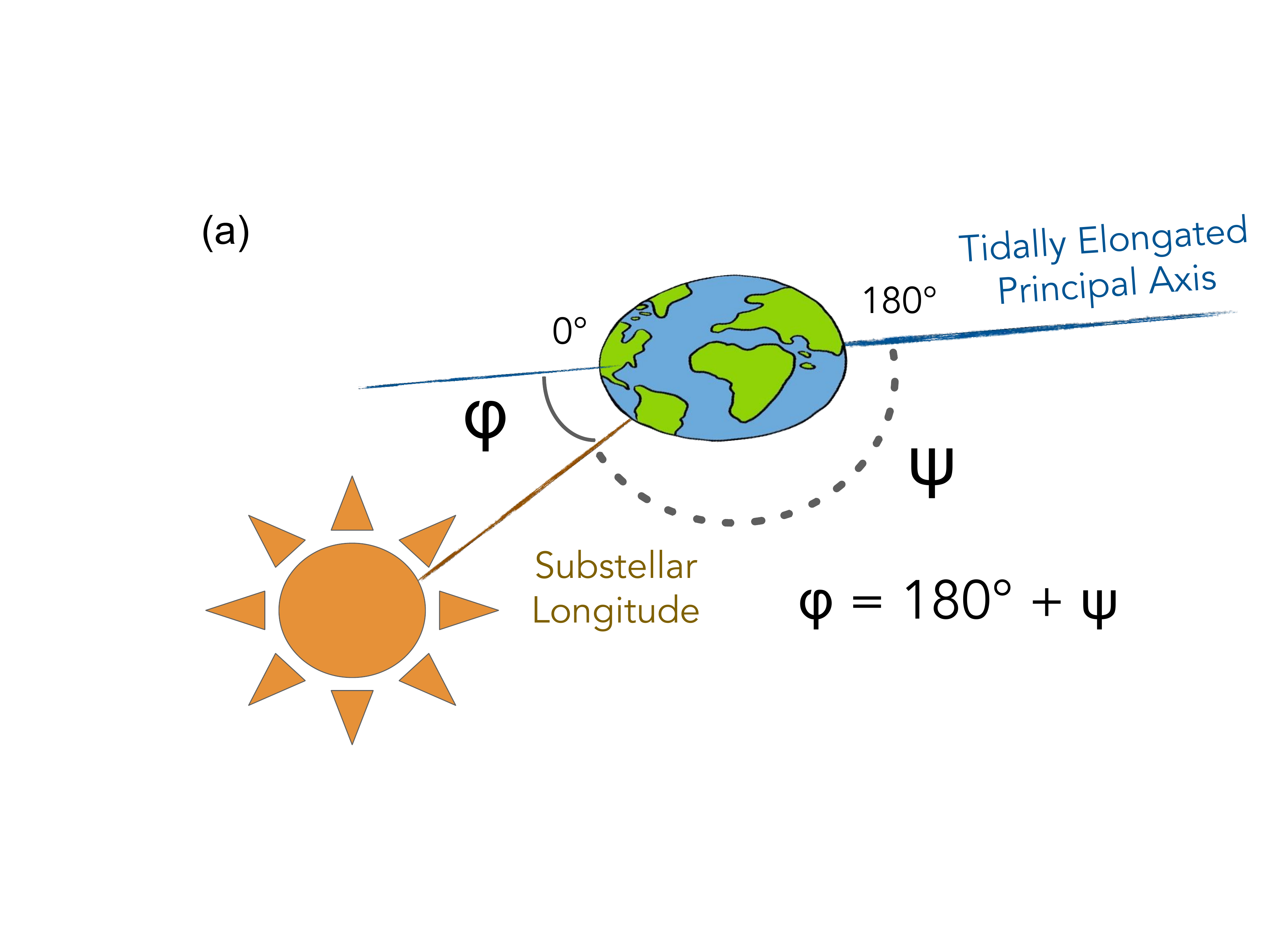}
\includegraphics[width=1.\columnwidth]{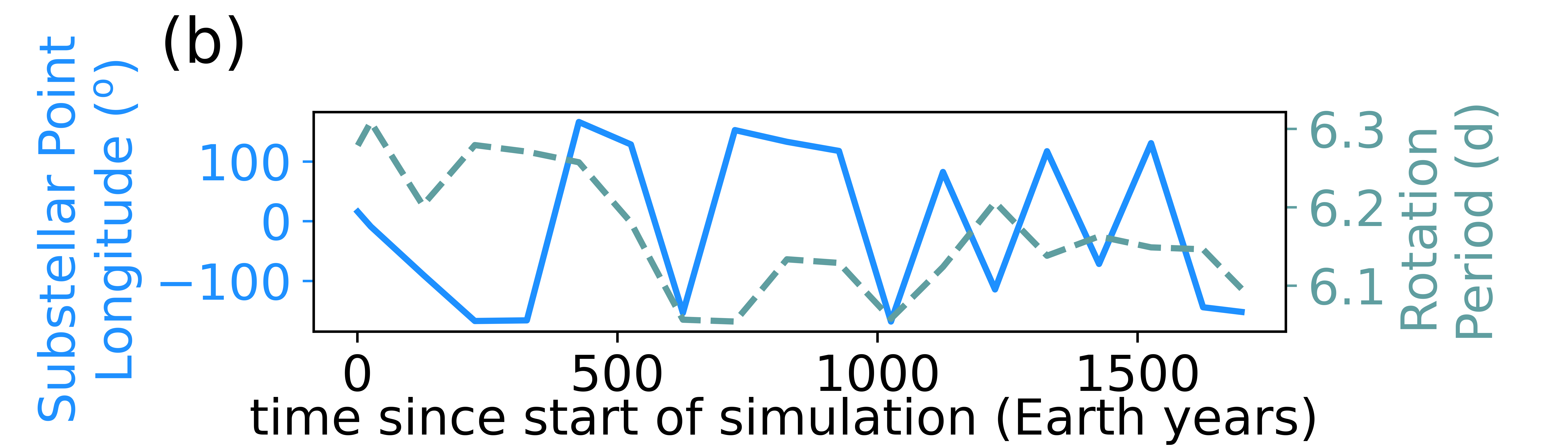}
\caption{\label{fig1}  Cartoon schematic illustrating the relationship between $\psi$, the long-axis misalignment and $\phi$, the substellar longitude (a), and example timeseries plot showing the variability of substellar longitude and planetary rotation period due to spin-axis dynamics calculated by an N-rigid-body integrator on a 100-yr timestep (b). This study explores the effects of variations in $\psi$ and $\phi$ on exoplanet climate using an intermediately-complex Earth System model. } 
\end{center}
\end{figure} 
\hfill \break

\section{Numerical Model, Data, \& Methodology}
\label{methods}

An N-Rigid-Body integrator with self-gravitating particles is used to model the effects of spin-orbit coupling. Informed by the outcomes derived from these results, a climate model of intermediate complexity is employed to simulate the planetary climates of TRAPPIST-1 e and f.

\subsection{GRIT: an N-Rigid-Body model to simulate the spin-orbit coupling of planetary systems}

We use an N-Rigid-Body simulation package, the Gravitationally interacting
Rigid-body InTegrator (GRIT; \citealt{Chen+21ApJ}), to compute the spin and orbital evolution of TRAPPIST-1 e and f. To account for spin-orbital dynamics and their mutual interactions, GRIT utilizes a Lie–Poisson algorthrm starting with first-principal
rigid-body dynamics. This integrator is an improvement upon previous seminal work \citep{Touma94} by way of relaxing the assumption of strictly near-Keplerian orbits, including any number of objects as rigid bodies in the system, as well as the inclusion of high-order implementations. Refer to \citet{Chen+21ApJ} for the detailed model breakdown.

GRIT is employed to generate the parameters relevant to the planet's orbital and spin evolution; these parameters include: orbital elements (e.g., semi-major axis, eccentricity, inclination) and spin parameters (e.g., spin rate, spin along the x-y-z axis, obiliquity, and the misalignment of the planetary tidally-elongated- or long-axis). The illustration of the misalignment of the long-axis is shown in Figure~\ref{fig1}, where a tidally locked planet will have near zero misalignment. Z-axis spin values are directly read as the planetary rotation rate and the axis misalignment  $\psi$ are translated to the longitude of incident stellar heating ($\phi$) via $\phi = 180 + \psi$. The sign of $\phi$ is calculated using a 3D rotational matrix centered on the z-axis.

We perform three sets of GRIT simulations: TRAPPIST-1e with no tidal dissipation, TRAPPIST-1f with no tidal dissipation, and TRAPPIST-1f with  tidal dissipation. Each of these is run for at least 5000 years and their solutions are recorded annually. 
We adopted a constant time lag model following \citet{Eggleton+98}, setting the time lag to be $2\e{-5}$ years (or 638 seconds, similar to
the case of the Earth). We set the Eggleton's $Q$ number ($Q_E$) to be 0.23,
corresponds to a love number of 0.3 (also similar to that of the
Earth).

\subsection{ExoPlaSim: a user-friendly and versatile GCM for exoplanet environments}

The Exo-Planet Simulator (hereafter ExoPlaSim) is an intermediately-complex GCM maintained and developed by \citet{Paradise+22}. ExoPlaSim has been modified and adapted for exoplanet environments and characteristics after the original Planet Simulator (PlaSim; \citealt{Fraedrich+2005}).

As GCMs such as ExoPlaSim simulate Earth-system components (i.e., the atmosphere, land, ocean, and sea-ice), they are advantageous over single-column or one-dimensional climate models. However, simplifications regarding certain parameterizations and processes such as radiative transfer or surface hydrology  are made \citep{Poulsen+2001,Vallis+2018,Paradise+19,Galuzzo+21}. Relaxing these assumptions with an intermediate-complexity GCM is appropriate as our study is not designed to reproduce climate behaviors with high realism nor predict specific exo-atmospheric compositions, but rather provide a more generally meaningful result. In addition, such GCMs with reduced complexity permit simulation of key 3D processes including cloud formation, larges-scale circulation, moist processes, and  climate feedbacks at a lower computational cost (see e.g., \citet{Biasiotti+2022} for another category of computationally inexpensive but accurate climate models). This will become crucial when we need to compute a diverse range of anticipated exoplanet compositions.

Here, ExoPlaSim is used to simulate Earth-like, CO$_2$-dominated, and steam atmospheres\footnote{Even though volatile accretion models suggest a diverse range of atmospheric compositions \citep{Chen+Jacobson22}, our goal is not to make any definitive predictions on them. Rather, these compositions are set to allow clement surface temperatures to arise (while avoiding the moist greenhouse threshold) to facilitate comparison between modeling results.} with planet properties (i.e., mass, radius, orbital period) consistent with TRAPPIST-1 e and f \citep{Agol+21}. Orbital eccentricity and precession are set to zero for all planets. For tidally locked planets, we set the rotation periods to be equal to their orbital periods and the substellar point fixed at 180$\degree$ longitude. For the perturbed planets, we update the rotation period and the substellar point of the planet every one Earth year, with spin values derived from GRIT outputs.  

Throughout the paper, we use the original 1:1 unperturbed substellar point ($180\degree$) as the reference point for ``current" location of the perturbed substellar point. The current substellar point longitude for any perturbed system is equivalent to $180\degree$ added or subtracted by the degree of long-axis mislignment ($\phi$) with the direction towards the star. The ``dayside hemisphere" refers to between 90 and +270\degree longitude, regardless of the actual location of stellar insolation.

\begin{figure*}[t] 
\begin{center}
\includegraphics[width=1.8\columnwidth]{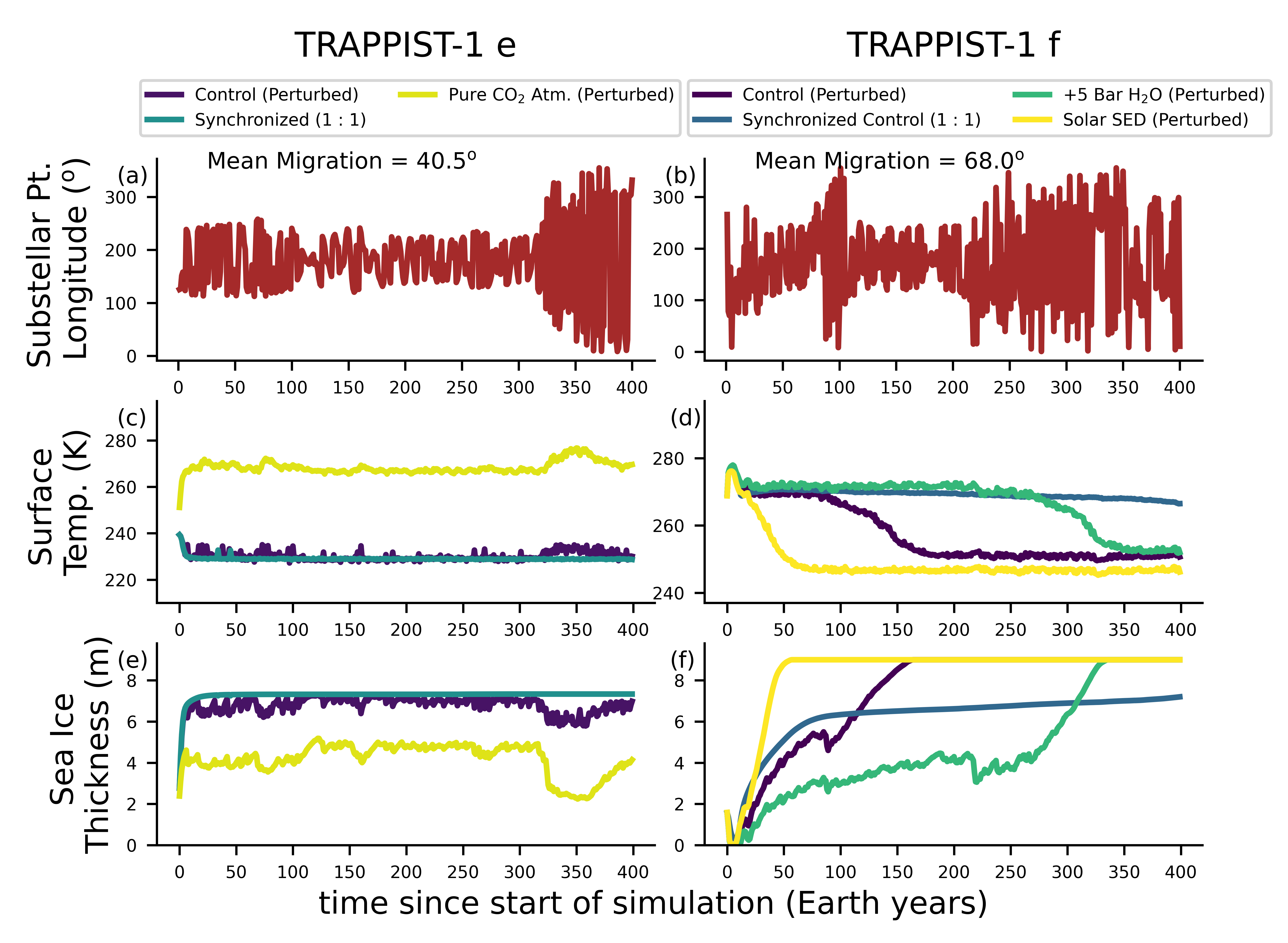}
\caption{\label{fig2} Timeseries of N-Rigid-Body simulation outcomes (a d) that are used as inputs to the climate model results for TRAPPIST-1 e and f. The top panels show the substellar point longitude  and the bottom panels show global-mean surface temperatures and sea-ice thickness. The inclusion of sporadic planet interactions does not affect the mean climate of planet e, while increase pace of global glaciation is found for planet f with the inclusion of planet interactions. In the latter planet scenarios, as the location of maximum stellar heating moves away from the open ocean areas, the high surface albedo of the newly formed sea-ice makes deglaciation extremely difficult, and the planet rapidly transitions into a snowball state over ${\sim}200$ years. 180\degree is the substellar longitude for the 1 : 1 resonant state.}
\end{center}
\end{figure*}

For planet e, we simulate three scenarios with  Earth-like atmospheric compositions:  aquaplanet (control), desert planet (i.e., 100\% land cover), and an aquaplanet with 1:1 synchronized orbits. For planet f, we simulate 33 bar CO$_2$ (control) and a 33 bar CO$_2$ plus 1 bar H$_2$O composition\footnote{Changing $p$H2O at the model configuration stage only changes the surface pressure and the mean molecular weight; the actual water vapor field is set entirely by moist processes such as evaporation, precipitation, and tracer transport through convection and circulation.}. The aquaplanets use 30-m thermodynamic slab ocean model and desert planets have constant surface albedo values of 0.2  \citep{Paradise+22}. The majority of the results will be focused on comparing the control runs with one or two of the sensitivity experiments. We also test the effects of two different stellar spectral energy distributions (SEDs).

For the Sun-like star SED, we use a reconstructed solar irradiance spectrum from \citet{lean1995reconstruction}. The input spectrum version is fixed in the year 1850 and no observed irradiance cycle is included. For the SED of the TRAPPIST-1-like star, we downloaded pre-generated spectra from the BT-Settl database (Husser et al. 2013). This spectra has stel lar metallicity of [F/H] = 0.0, alpha- enhancement of [$\alpha$/M] = 0.0, surface gravity log g = 4.0, and stellar effective temperature ($T_{\rm eff}$) of 2600 K.

All ExoPlaSim simulations use an exponential filter, which is applied both at the transform from gridpoint space to spectral space, and then from spectral space back to gridpoint space. Appendix A of  \citet{Paradise+22} state that at T21 and T42 resolutions, exponential and Lander-Hoskins filters have very similar performance, with the former working better for slower rotators ($> 7$ days). We set the T21 horizontal resolutions with vertical domains extending up to 0.01 bar. All simulations are integrated for at minimum 300 Earth years and maximum of 800 Earth years (hereafter, years will refer to Earth years). We consider the radiative balance to be at equilibrium after the 30 year mark.

\begin{figure*}[t] 
\begin{center}
\includegraphics[width=1.8\columnwidth]{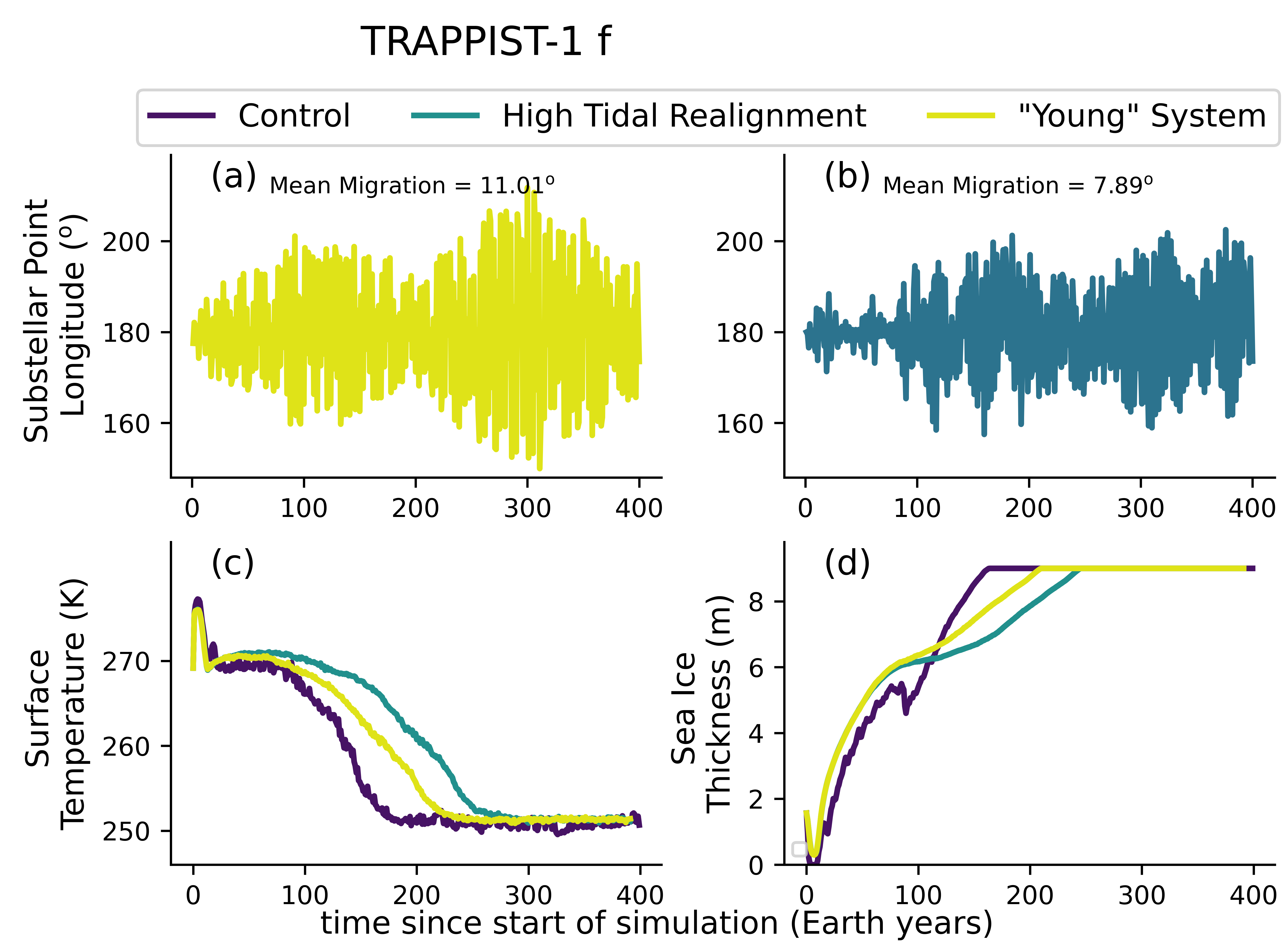}
\caption{\label{fig3} Timeseries of N-Rigid-Body simulation outcomes (a, b) that are used as inputs to the climate model results for TRAPPIST-1 f (c, d), exploring different levels of system excitation. The top panels show the substellar point longitude  and the bottom panels show global-mean surface temperatures and sea-ice thickness.  }
\end{center}
\end{figure*}

\begin{figure*}[t] 
\begin{center}
\includegraphics[width=1.7\columnwidth]{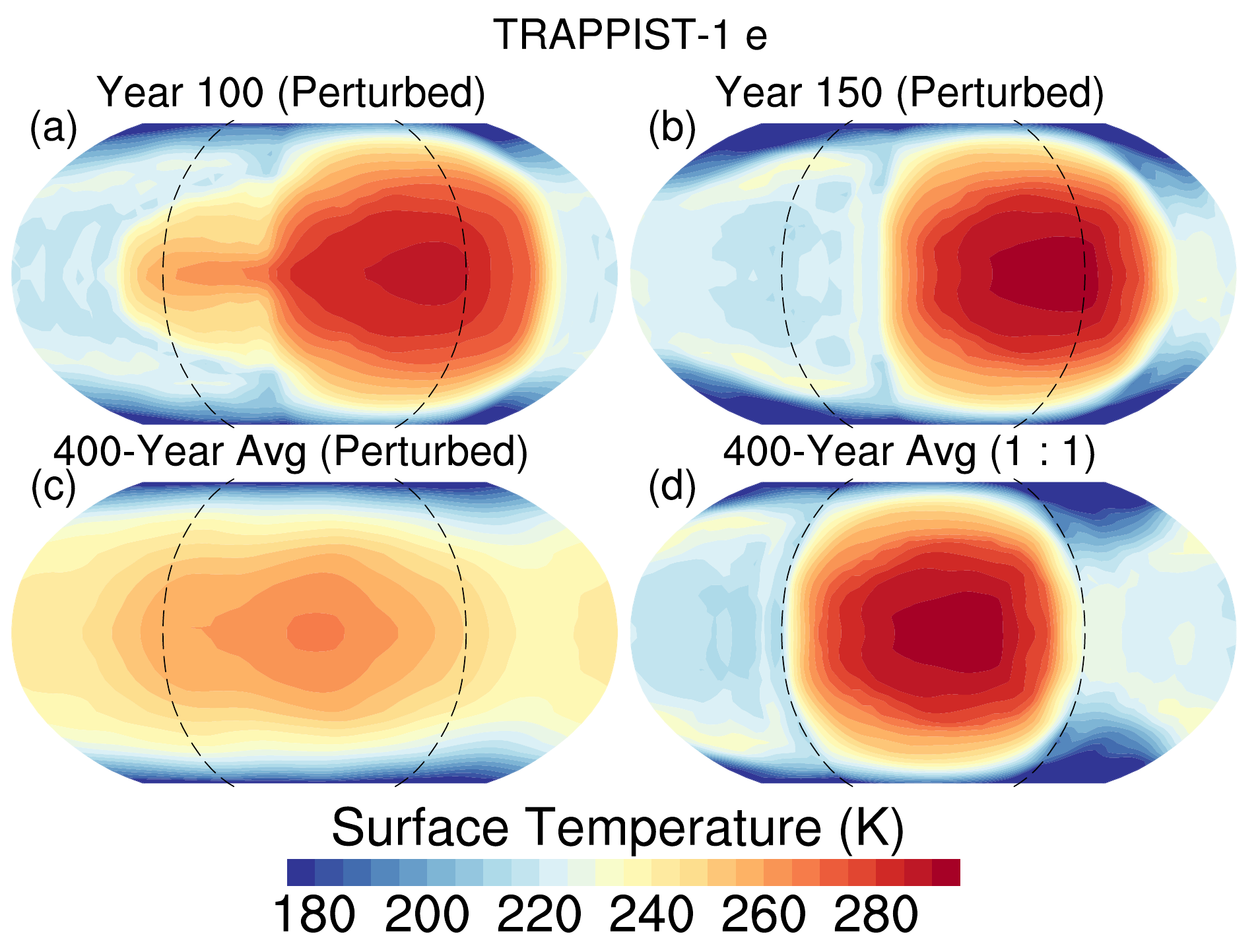}
\includegraphics[width=1.7\columnwidth]{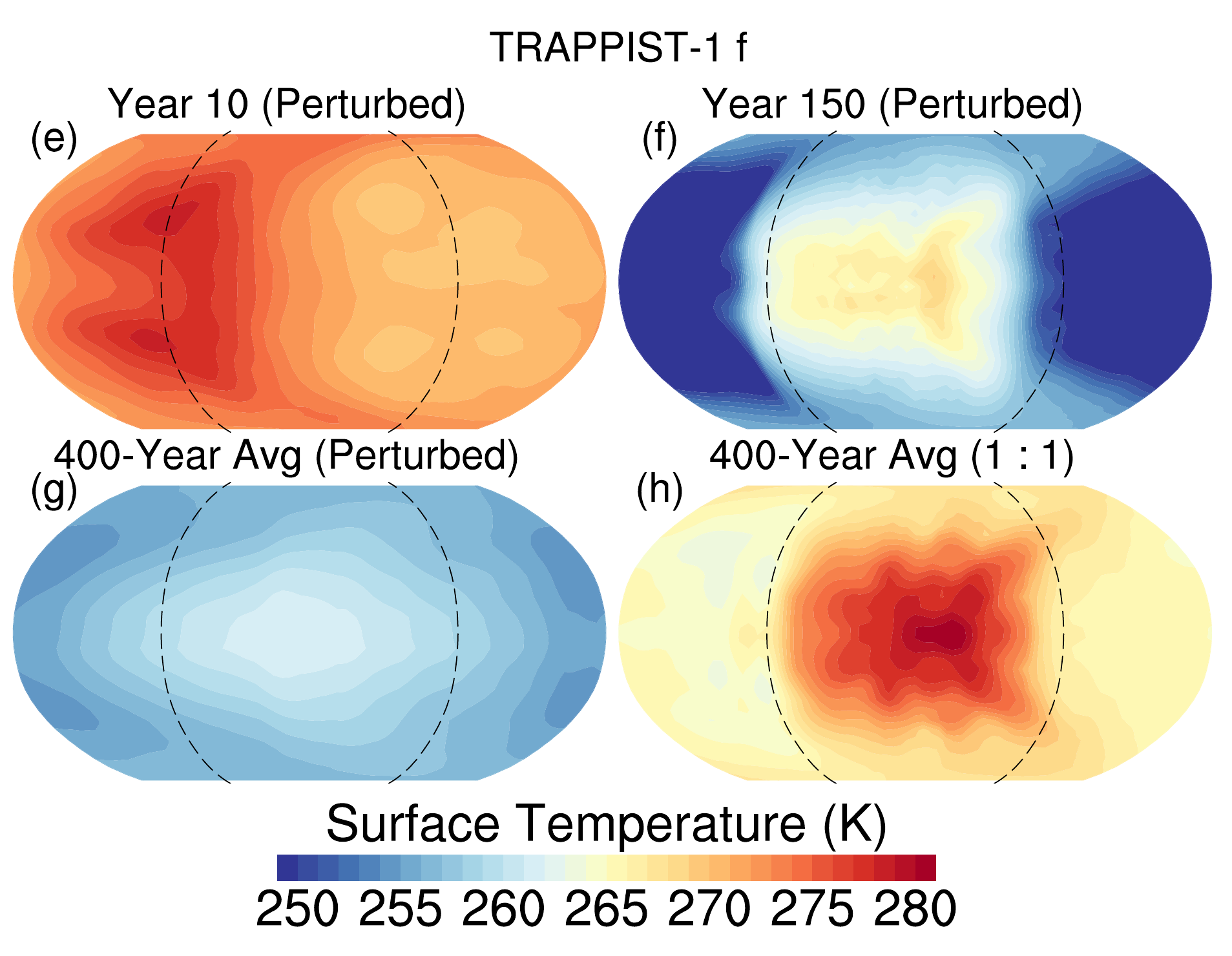}
\caption{\label{fig4} Map projections of sea surface temperature for perturbed planet e (year 1, year 100, and 400-year averaged; a, b, c respectively) and unperturbed planet e (d) as well as those for perturbed planet f (year 1, year 100, and 400-year averaged; e, f, g respectively) and unperturbed planet f (h).  We find that time averaged climates of perturbed scenarios deviate from a fixed 1:1 tidally locked scenario. The spatially and temporally varying temperature and albedos may imprint themselves in observations e.g., thermal emission measurements. An animated adaptation of this figure is available in the online publication. The animated version shows the sea surface temperature evolution of planet f annually from 0-400 years (left) and the time-averaged climate for the cumulative time period (right). }
\end{center}
\end{figure*}

\subsection{Leveraging N-Rigid-Body Outputs in the 3D Climate Model: Some Details and Caveats}
 The climate model was used to simulate a 400-year time slice from the full 5000-year GRIT output. TRAPPIST-1 e control and TRAPPIST-1 f used  4600 to  5000-years to examine the later stages of orbital evolution when the systems have been sufficiently excited.  The ``younger" planet f used the same GRIT output as the control, but started at time zero; the simulation with high tidal dissipation also started at time zero. Changes to substellar forcing informed by GRIT results are applied as soon as each climate model run starts, and these changes are updated every year according to these solutions. 
 
Effects due to eccentricity variations and precession, while not explicitly calculated, are reflected in our spin-misalignment calculation. Specifically, $\psi$ is the long-axis misalignment with respect to the direction towards the star, and the direction towards the star is calculated based on the orbital location of the planet, which is affected by eccentricity and orbital precession. We do not include obliquity variations in this study since the planets do not acquire large obliquities (the maximum obliquity of TRAPPIST-1 e is $5\degree$, and that of f is $1\degree$) in the current simulation set. Obliquity can be larger depending on the initial configurations, but we do not consider this in order to isolated the effects due to spin-axis misalignment.

One advantage of our modeling framework is the ability to consider the effects of spin dynamics on orbital evolution. In our numerical experiments of the TRAPPIST-1 planets as point-mass particles, this has shown to be relatively weak due to greater angular momentum of the orbit relative to planetary spin.  However, this effect might be important for other planetary systems.

\begin{figure*}[t] 
\begin{center}
\includegraphics[width=1.7\columnwidth]{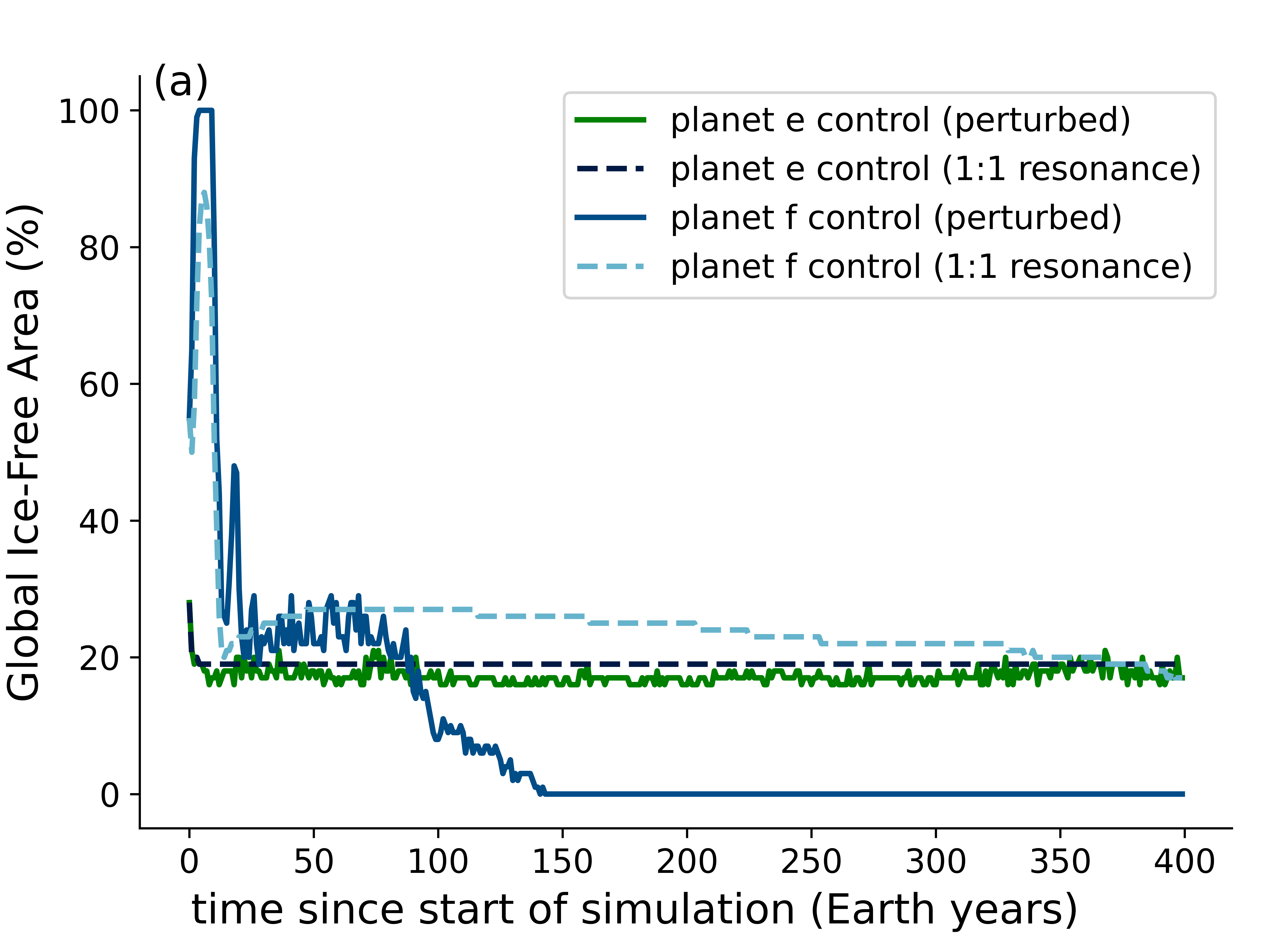}
\includegraphics[width=1.8\columnwidth]{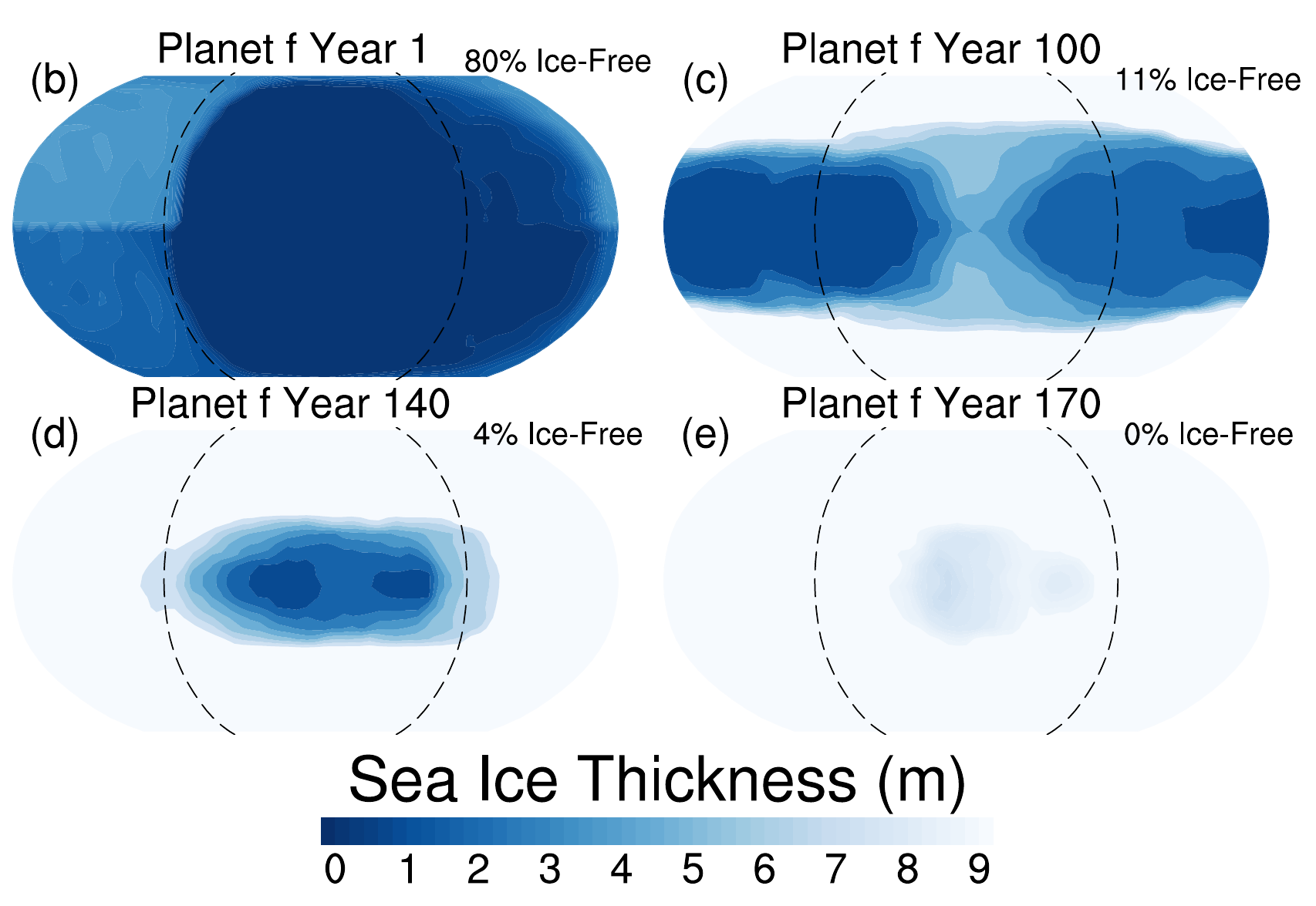}
\caption{\label{fig5} Temporal evolution of global ice-free fraction (a) and yearly-averaged map projections of sea ice thickness for planet f control at various epochs  (b, c, d, e). Formation of sea ice is determined by the sea surface temperature. An animated adaptation of this figure is available in the online publication. The animated version shows the sea-ice thickness evolution of planet f annually from 0-400 years (left) and the time-averaged sea-ice thickness for the cumulative time period (right).  }
\end{center}
\end{figure*} 

\hfill \break

\section{Results}
\label{results}

Our results show that the libration of the substellar hemisphere due to mutual gravitional interactions between neighboring planets can substantially alter exoplanet climates at the outer edge of the habitable zone. We demonstrate this by comparing N-Rigid-body simulation and climate model results of TRAPPIST-1 e and TRAPPIST-1 f. Synchronized planets (i.e., 1:1 resonance) will be hereafter referred to as unperturbed planets, whereas non-synchronized planets (i.e., under the influence of planet interactions) will be referred to as perturbed planets.

For both TRAPPIST-1e and f, planet-planet interactions lead to non-stationary substellar longitudes. For planet e, the migration remains small (${\sim}40\degree$ longitude)  until the system is sufficiently excited after year 330 (Figure~\ref{fig2}a).
On the contrary, for planet f, we find that the substellar point rapidly deviates from the synchronized state, and even at times passes through the anti-stellar position (near $360\degree$ longitude; Figure~\ref{fig2}b). Only in certain time intervals (e.g., between 130 and 200 years) do the substellar  point linger around the original position. Comparing the substellar longitudinal evolution of planets e and f, the former has greater symmetry between the eastern and western hemispheres.  

For planets e and f, the most obvious feature of perturbed planets is the fluctuations in global-mean surface temperature and sea-ice thickness (Figure~\ref{fig2}c-f), whereas the unperturbed planets have smoother temperature and sea-ice curves. For planet e,  the $T_s$ difference between the perturbed and unperturbed is small, even when substantial substellar point migration has been introduced at the later stages (Figure~\ref{fig2}c). This can be seen by their partially overlapping temperature and sea-ice thickness curves. However, for planet f, the exact composition of the atmosphere or whether planet-planet interactions were present dramatically affected its evolutionary pathway. For instance, the $T_s$ of the pure CO$_2$ atmosphere (control) becomes fully  glaciated at year 175, whereas the CO$_2$ + H$_2$O atmosphere enters that state at year 300 (Figure~\ref{fig2}d). When the rate of sea-ice formation of the perturbed control outpaces that of the unperturbed at year 100 (Figure~\ref{fig2}f), the location of maximum stellar heating moves away from the open ocean areas and the resultant high surface albedo of the newly formed sea-ice makes deglaciation extremely difficult.

\begin{figure*}[t]            
\begin{center}             
\includegraphics[width=2.1\columnwidth]{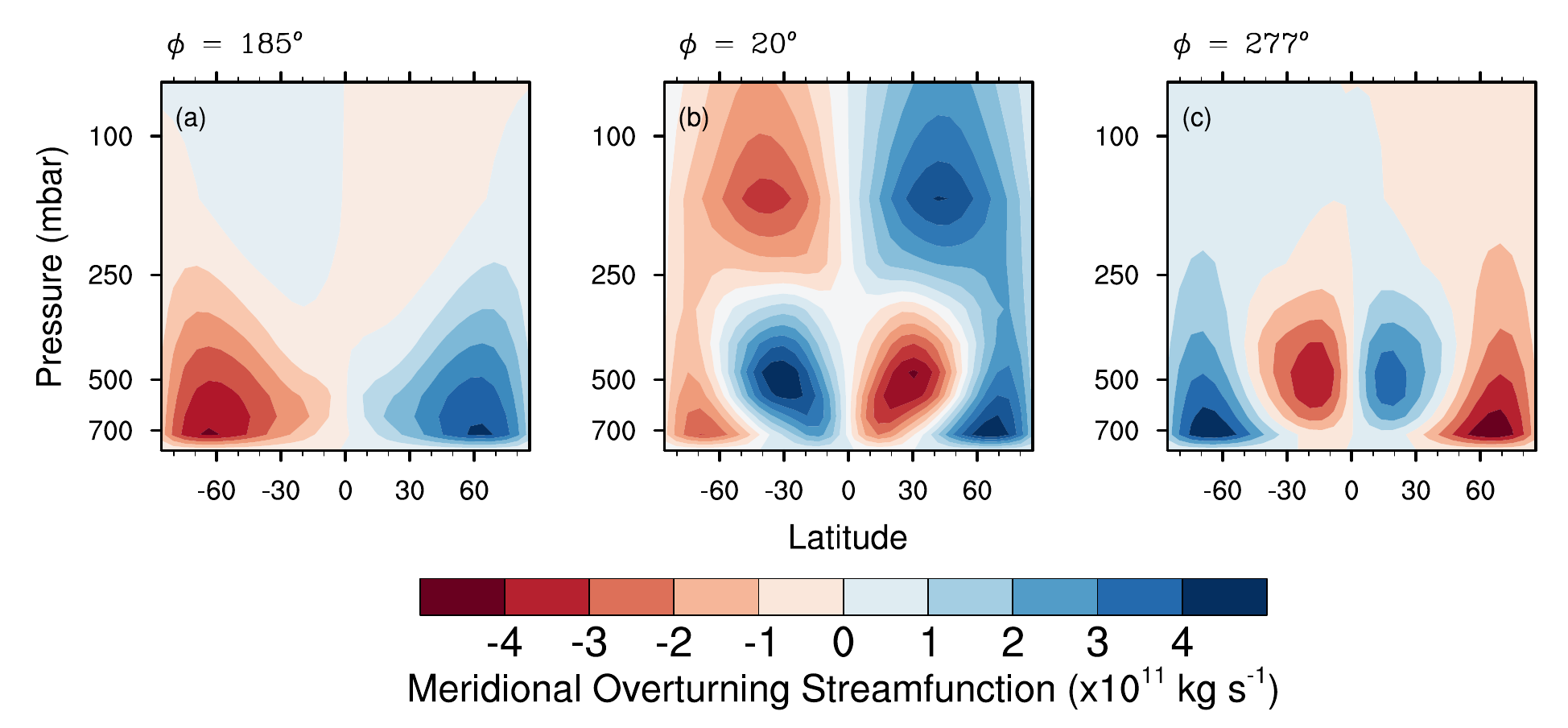}
\caption{\label{fig6} Hemispheric-mean mass stream functions at three snapshots in time for TRAPPIST-1 e with Earth-like atmospheres, averaged between longitude 90\degree and 270\degree. We show that the patterns of the overturning circulation can be modified dramatically depending on the exact longitudinal location of  stellar heating: a) 185\degree, the original position, b) 20\degree, near the antistellar point, and c) 277\degree, at the terminator. The first scenario represents the 1:1 state, the second indicates downwelling at the ITCZ, and the third is an example of a ``transitional" circulation pattern. }
\end{center}
\end{figure*}


A planetary system can exist in lower excited states than what has been discussed in Figure~\ref{fig2}; whether these states would result in substantial sea ice build up deserves further investigation. Here, we find that for a younger system, the deviation of the substellar point from the 1 : 1 substellar longitude is only about 15$\degree$ (Figure~\ref{fig3}a). However, the degree of libration is most suppressed when the stellar tidal force from the host is elevated (Figure~\ref{fig3}b).  In this scenario, the first 80 years have very synchronized orbits. Even after 100 years, the average deviation is only 10$\degree$ while never exceeding 20$\degree$ throughout the full timeseries. In all three scenarios, the onset of runaway glaciation still occurs prior to the unperturbed with a 20 year delay of the global glaciation timing\footnote{While these intervals seem short, the difference between the climate evolution tracks of each planet is likely much greater due to the high thermal inertia of more realistic ocean basins.} between each simulation (Figure~\ref{fig3}c, d).

Figure~\ref{fig4} show snapshots of surface temperatures averaged across various epochs for the control runs for each planet. Temperature maxima generally indicates the location of stellar heating, and when the shift in the location of substellar heating is rapid, remnant heat imprints can be preserved  (Figure~\ref{fig4}a and Figure~\ref{fig4}e). On average, the $T_s$ projections of planets e and f show much shallower day-night temperature gradients in comparison to that of the unperturbed planets (Figure~\ref{fig4}c and Figure~\ref{fig4}d). The 400-year mean of planet mirrors the transient/annual averages in terms of the relative location of stellar heating and $T_s$ distribution (implying that climate equilibrium is quickly established with each change in the substellar longitude). For planet f however,  the temperature distributions at all three epochs are vastly different compared to that of the synchronized, reflecting the much greater influence of libration on climate in this regime. Overall, the most notable difference between planets e and f is the greater day-night and equator-to-pole temperature contrasts. The much thicker atmospheres needed to maintain open oceans for planet f (33 bar versus 1 bar of planet e) lead to more efficiency heat transport and thus muted temperature contrasts.

A consequence of spin-orbit modulated $T_s$ is the effect on sea-ice formation. Of the four scenarios shown, the only one devoid of any ice-free regions at year 400 is planet f control, which rapidly glaciates from the start, slowly glaciates after ${\sim}30$ years, and then completely glaciates by the 140 year mark (Figure~\ref{fig5}a). Ice-free fraction of planet f in the 1:1 state quickly decreases in the first 30 years, slightly increases, and then returns to about 30\% at year 300.   Planet e experiences higher instellation and lower degrees of substellar migration, leading to minimal differences in ice-free area fractions between the perturbed and unperturbed planet.

Figure~\ref{fig5}b-e shows snapshots of sea-ice thickness for planet f control. As can be seen, the planet began with a full ocean cover (b), and as the the substellar point migrates away from the original location, the sea-ice closes in towards the equator, dramatically elevating the surface albedo and making the planet more difficult to warm via the ice-albedo effect. At year 140, remnant open ocean region is concentrated in the  region of maximum stellar heating, leading to a familiar eyeball pattern. This pattern persists for 30-40 years before eventually transitioning into a global snowball state (d).  
The difference in the above between planets e and f are largely due to the thicker atmospheres, lesser incident flux, and greater libration degrees of planet f. From the standpoint of the initial substellar hemisphere centered at 180\degree, it begins evolving from the warm state and as the stellar flux is shifted longitudinally, the regime transitions into a glaciated state through rapid sea ice build up. When the substellar hemispheres librates back and the stellar flux here returns to the earlier value, the surface remains in a frozen state instead of the previous warm state it enjoyed, and would  require much greater stellar fluxes to deglaciate the planet. Conversely, none of the planet e cases display such a behavior as the higher stellar fluxes and thinner atmospheres removes the possibility of substantial climate bifurcation. With Earth-like atmospheric pressures, all planet e scenarios result in eyeball states (Figure~\ref{fig4}d). Stronger instellation and reduced rate of substellar point migration allows these cases to sustain open oceans for much longer periods compared to those of planet f.

In addition to climate forcings, here we provide a glimpse of the effects of spin-orbit variations  on atmospheric dynamics.
As the planet librates longitudinally over ${\sim}100$ years, the zonally averaged flow across the entire globe is not substantially disrupted. However,  between 90 and 270\degree longitude, massive shifts in the mass streamfunctions can be seen by leveraging the Eulerian-mean meridional velocity and the generalized vertical coordinates (Figure~\ref{fig6}). In the circulation regime of TRAPPIST-1 e when the substellar longitude is at 185\degree, eddy momentum and heat sources are minimal, and thus the differential diabatic heating is balanced only by adiabatic cooling (latent heating) near the substellar point and adiabatic warming (radiative cooling) at higher latitudes, resulting in extended single cells in each hemisphere with strengths of ${\sim}10^{11}$ kg s$^{-1}$. This situation does not hold  when the substellar point shifts to 20\degree (Figure~\ref{fig6}b).  At this near ``antistellar" position (relative to the initial substellar hemisphere), the geometry of the meridional overturning circulation starts to reverse to a substellar downwelling and high latitude upwelling pattern. At 277\degree, the mean-flow transitions to  mixtures between stronger equator to mid latitude cells and weaker cell near the poles (Figure~\ref{fig6}c). Overall, these results highlight unique circulation features on librating planets may have strong implications for day-to-nightside transport of dust, aerosols, and gas-phase species \citep{Boutle+2020,CaroneEt2018MNRAS}. Due to rapid homogenization of diabatic heating induced dynamics that mimic the mean-flow of Earth-like rotators \citep{Turbet+2018AA}, asynchronous worlds in higher SORs would not show such behavior in atmospheric dynamics and their variability.
\section{Discussion}
\label{discussion}

To date, the majority of climate modeling of rocky HZ exoplanets have focused on synchronized planets and those with higher order spin-orbit resonances (e.g., 3:2, 2:1, 6:1), due to the assumed likelihood that most planets we have discovered will be in these states. Our results show that stochastic libration caused by secular gravitational interactions could modulate the climates of exoplanets in systems such as TRAPPIST-1\footnote{TRAPPIST-1 is found to be highly active in the X-ray and extreme ultraviolet, rendering its planets vulnerable to atmospheric erosion and making our atmospheric composition/pressure assumptions debatable ($\ga 20$ bars) \citep{Roettenbacher+17,Seli+21,Krissansen+22,BeckerEt2020AJ}.  However, the planetary parameters taken here were only used as a proof of concept to demonstrate climate variability in general  and our goal was not to  make conclusions regarding the habitability of any particular world/system.}.

The orbital and spin trajectories of planets in compact systems have been investigated in several studies using N-body/particle simulations (e.g., \citealt{Rice+18,Vinson+2019,Satyal+22}).
However, GRIT has a number of significant advantages over previous work. The majority of previous models (e.g., REBOUND, SymBA) assume that the planetary bodies or other objects are point-mass particles, and then spin-axis dynamics are calculated after-the-fact. Thus, effects of planetary spin on their orbital dynamics are neglected. In addition, previous studies typically included a 1-dimensional study on the spin evolution, and changes in obliquity (spin-orbit misalignment) were neglected \citep[e.g.,][]{Vinson+2019,Brasser+22}. The inclusion of  obliquity (effects in the z-direction) mediates and therefore reducing the degree of libration. 
GRIT, which is based on the first principle rigid body dynamics, provides more accurate results on the spin and orbital evolution.  Future work will include a more detailed study on the statistical properties of spin variations based on different planetary tidal parameters, and investigate the effects on climate and photochemistry.

Many authors have previously argued that planets at the OHZ may be better candidates for biosignature and habitability-indicator targets as they are less susceptible to atmospheric escape and irreversible water-loss \citep{Rogers+2021, LammerEt2011A&A}, in comparison to those at the IHZ. Further, \citet{Colose+2021} found that tidally heated planets with eccentric asynchronous orbits around cool stars may be able to maintain temperate climates even with stellar fluxes of ${\sim}600$ W m$^{-2}$ (Mars-like). However, the effects of tidal heating will likely be diminished for planets at the OHZ \citep{Dobos+19}, corresponding to those experiencing the largest degrees of libration in this work (i.e., change in 50$\degree$ longitude in ${\sim}100$ years on average; Figure~\ref{fig1}). 
In compact multiplanet systems, we argue that OHZ planets should rapidly transition into snowball states with modest-to-low stellar and tidal heating. When the stellar heating returns to its original position, the planet still remains in a snowball state. 
For temperate planets (those with fluxes between 900 and 1400 W m$^{-2}$), we agree with previous studies that they are the most promising targets for future atmospheric characterization campaigns. However, for single-planet systems or those without resonant chains, the spin-axis dynamical effects studied here do not come into play, hence the conclusions cited above regarding the habitability of those planets are unchanged.



The most significant difference between our results and those using more complex GCMs (e.g., ExoCAM, ROCKE-3D, LMD, UM) is the amount of greenhouse gas needed to substantially warm up surface air temperatures, especially in colder regimes of the HZ. Compared to \citet{FauchezEt2019ApJ}, our simulations require ${\sim}10$ bars or more of CO$_2$ to substantially warm TRAPPIST-1 f (the results for TRAPPIST-1 e are the same). One likely explanation for this discrepancy is that ExoPlaSim and PlaSim heritage models place clouds at the substellar point of tidally-locked models \citep{CheclairEt2019ApJL,Paradise+19}. Yet more-advanced cloud models in  state-of-the-art GCMs have shown that the daysides of tidally-locked planets receiving lower stellar fluxes might be only partially cloudy. For example, the clear substellar skies found in GCM simulations typically have substellar point TOA albedo of ${\sim}0.05$, whereas ExoPlaSim's substellar point has a TOA albedo of  0.8-0.9 (since clouds are assumed to be gray). Different albedos would lead to differences in total energy budget and therefore surface temperatures. Gaps in physical realism, for instance in modeling sea ice and convective processes, might also affect these predictions. Hence in certain regimes e.g., the IHZ and OHZ where water cloud formation and condensation respectively become important, one would require verification or complementary experiments with more complex GCMs and with different radiative schemes. In any case, further model comparison efforts are needed to clarify these assumptions, parameterizations, and model uncertainties (e.g., \citealt{YangEt2019ApJ,Fauchez+21,Haqq-Mirsa+22}).

In addition to climatic influences, those related to photochemistry and the formation of clouds and hazes may be markedly different if alternating regions of the planet are exposed to the star. For instance, the impact of stellar UV flare events
\citep{howard+2022b,Paudel+21,Louca+2022arXiv} and stellar plasma (e.g., protons and $\alpha$-particles; \citealt{Chen+2021}) on varying sides of the hemisphere may lead to different global chemical and participle precipitation rates. Moreover, distinct atmospheric dynamics due to oscillating planetary spin (Figure~\ref{fig5}) could drive the planet into other  transport regimes \citep{CaroneEt2018MNRAS,ChenEt2019ApJ} and change how chemical species and aerosols are advected from the dayside to the nightside \citep{Boutle+2020,Cohen+22MNRAS}. Modulations in vertical mixing region and strength would lead to altered photolysis rates and shallower flux-abundance curves of biosignature gases.

Future work should also include the use of a fully coupled GCM with a dynamic ocean. \citet{Way+2017} studied the effects of variable eccentricity of Earth-like planets in the absence of Mars and found that relative humidity, precipitation, and sea ice fraction vary on the order of ${\sim}5,000$ yrs. Others have demonstrated how GCM simulations with the inclusion of ocean circulation and salinity can depart considerably from those established by slab ocean models. In addition to equilibrating on much longer timescales (500-1000 yrs), coupled ocean model found increase regions of surface liquid water extending even to the nightsides due to the effects of ocean heat transport \citep{Hu+Yang2014PNAS,DelGenioEt2019AsBio,Olson+20,SalazarEt2020ApJL}. At the OHZ, broadened open ocean basins would help delay the onset of the snowball states for planets under the influence of librating substellar hemispheres.   Additionally, previous work ruled out the possibility of limit-cycles on tidally-locked M-dwarf planets   due to the continuum of equilibrium states at all sea ice fractions \citep{CheclairEt2017ApJ,CheclairEt2019ApJL}. Our study suggests that hysteresis in this regime is still possible. However, the degree of hysteresis is  likely  small and may not necessarily entail the recovery of geochemical limit cycles driven by outgassing and carbon-silicate weathering. Further investigation with coupled ocean models is warranted  to firmly establish the role hysteresis in this regime.

Finally, the conclusion reached here are not strictly applicable to  M-dwarf worlds experiencing MMRs. A variety of strong planetary interactions can also lead to spin variations and dynamically-interesting  atmospheres. In the absence of these processes, one might assume that the planets in question would simply fall back to synchronized states\footnote{Alternatively, if the planet has high eccentricity ($e>0.2$), \citet{Renaud+21} showed that higher order SOR configurations are the most plausible.}. However, the complete absence of external perturbers may be rare occurrences, as evidenced by the plethora of satellites in the solar system and the high likelihood for the prevalence of (exo-)moons from recent surveys \citep{Teachey+18}. Further, many compact systems exists for multiple planets with K star hosts, e.g., Kepler-411 and K2-266, \citep{Rodriguez+18,Sun+19}, suggesting that our implications for the OHZ are relevant for planetary systems beyond those that are TRAPPIST-1-like, and even extend to planets born with a variety of rotation rates around Sun-like stars \citep{Kane+19AJ,Guzewich+20}. Binary systems may also be sites for such perturbations, but the stability of their planetary orbits is debatable (see e.g., \citealt{Quarles+22,Forgan16}). These possibilities will need to be examined with different star-planet-disc boundary conditions such as initial dynamical frictions in order to assess the competition between the strength of host star tidal realignment and external interactions with other rocky bodies.

\hfill \break
\section{Conclusion}
\label{conclusion}

Exoplanets residing in multi-planet compact systems are often assumed to be lodged in  1:1, 3:2, and 2:1 resonant chains. However, sporadic and highly-variable planetary spins and orbits may have drastic climate and atmospheric consequences. Here, for the first time, an N-Rigid-Body spin-orbital integrator is used in conjunction with a general circulation model to investigate the climate of TRAPPIST-1 e and f under the influence of host star tides and planet-planet interactions.
We find that the secular gravitational interactions between mutual orbits in compact systems can drive planets out of synchronized states, affecting their evolving and mean climates. This effect is particularly dramatic for planets further away from the host star, due to reduced strength of tidal dissipation. We further find that it is challenging to sufficiently warm planets at the outer edge of the habitable (even with CO$_2$-rich atmospheres in excess of 40 bars) due to greater degrees of substellar longitude migration and increased climate hysteresis. As this drift occurs on decadal timescales, it allows the formation of new sea-ice which increases the surface albedo of the planet, making subsequent deglaciation by stellar heating difficult. Our study suggests that OHZ planets in compact systems are less likely to have significant regions of open ocean basins, even with $>1-10$ bars of greenhouse gas warming.

Temporally and spatially variable temperature contrasts between the day and nightsides of these planets could manifest themselves in secondary eclipse thermal emission spectra. 
Moving forward, we will employ climate models in conjunction with N-body or N-Rigid-Body simulations to systematically investigate the effects of planetary orbit and spin characteristics on exoplanet climate and atmospheric observables. 

\acknowledgements
Goddard affiliates acknowledge support from the GSFC Sellers Exoplanet Environments Collaboration (SEEC), which is supported by NASA's Planetary Science Division's Research Program. H.C. is supported by an appointment to the NASA Postdoctoral Program at Goddard Space Flight Center, administered by Oak Ridge Associated Universities under contract with NASA.  G.L. is grateful for the support by NASA 80NSSC20K0641 and 80NSSC20K0522. We acknowledge high-performance computing support from Cheyenne (doi:10.5065/D6RX99HX) provided by NCAR's Computational and Information Systems Laboratory, sponsored by the National Science Foundation. This work used the Hive cluster, which is supported by the National Science Foundation under grant number 1828187.

\software{GRIT \citep{Chen+21ApJ}; 
ExoPlaSim \citep{Paradise+22}. Upon publication, the code for the climate model, input data, and scripts to
post-process and visualize the results will be available on GitHub: \url{https://github.com/hwchen1996/sporadic_climate}.  }

\bibliographystyle{apj}

\end{document}